# Microstuctural Changes Influencing the Magnetoresistive Behavior of Bulk Nanocrystalline Materials

**Stefan Wurster [1,*], Martin Stückler [1], Lukas Weissitsch [1], Timo Müller [2] and Andrea Bachmaier [1]**

[1] Erich Schmid Institute of Materials Science of the Austrian Academy of Sciences, Jahnstrasse 12, 8700 Leoben, Austria;

[2] Deutsches Elektronen-Synchrotron (DESY), Photon Science, 22607 Hamburg, Germany; Current Address: Anton Paar GmbH, 8054 Graz, Austria

***** Correspondence: stefan.wurster@oeaw.ac.at

**Abstract**

Bulk nanocrystalline materials of small and medium ferromagnetic content were produced using severe plastic deformation by high-pressure torsion at room temperature. Giant magnetoresistive behavior was found for as-deformed materials, which was further improved by adjusting the microstructure with thermal treatments. The adequate range of annealing temperatures was assessed with in-situ synchrotron diffraction measurements. Thermally treated Cu–Co materials show larger giant magnetoresistance after annealing for 1 h at 300 °C, while for Cu-Fe this annealing temperature is too high and decreases the magnetoresistive properties. The improvement of magnetoresistivity by thermal treatments is discussed with respect to the microstructural evolution as observed by electron microscopy and ex-situ synchrotron diffraction measurements.

## 1. Introduction

While the change in resistance in magnetic fields of ferromagnetic materials is small but well known for more than 150 years [1], the groups of Grünberg et al. and Fert et al. showed that changes of resistance with magnetic field, especially at low temperatures, can be gigantic when using thin layers of alternating Fe and Cr [2,3]. Some years after the discovery of the giant magnetoresistance (GMR) for layered thin film systems, it was also found for thin films and for bulk materials containing finely dispersed small ferromagnetic particles (granules) [4–8]. Thus, the effect was labelled granular GMR. Materials processed by severe plastic deformation (SPD) using high-pressure torsion (HPT) show such a GMR effect [9–11] and this group of materials is in the focus of this investigation. As a definition, GMR can be distinguished from ordinary magnetoresistance (OMR) and anisotropic magnetoresistance (AMR) by its negativity in all directions of applied magnetic field with respect to the current flow direction [12] and furthermore by its isotropic behavior [5]. This gives the evident advantage of 3D-bulk materials over 2D-thin film materials in the aspect of shaping when going towards applications.

Within this publication, GMR will be calculated according to

$$GMR = \frac{\Delta R}{R} = \frac{R(H) - R(H=0)}{R(H=0)} \quad (1)$$

taking into account the resistance R within the magnetic field H and at zero field. Although the present study focusses on the GMR behavior of nanostructured bulk materials, many important observations were found in studies on thin films. Wang and Xiao [13] investigated different types of binary systems, including Co–Ag, Fe–Ag, Fe–Cu, Fe–Au, and Fe–Pt. Thin film samples were produced using magnetron co-sputtering and magneto-transport properties in the as-deposited and annealed states were measured. The main findings of this study can be summarized as follows:

1. The (partial) immiscibility of both phases is a prerequisite for the appearance of granular GMR. The solid solution forming Fe–Pt did not show GMR behavior. Regarding the systems in focus of this publication, the solubility of Cu and Co and vice versa is smaller than 1 at% at 900 K [14]. The same is true for Cu–Fe below ~1000 K [15].
2. A peak in GMR occurs for ferromagnetic volume fractions between 15% and 25%. This was found to be universal for all systems showing granular GMR. Within the Cu–Fe-series, the highest GMR was found for $Fe_{20}Cu_{80}$ (in vol%).
3. The strength of GMR is affected by subsequent thermal treatments of the magnetron sputtered material. GMR is slightly larger for the Fe–Ag-system and for the Co–Ag-system after thermal treatment between 200 °C to 330 °C (depending on the chosen system) for 15 min and decreases again for even higher temperatures. For the Fe–Cu system, the maximum value was already found in the as-deposited state.
4. A reduction of the saturation field for GMR is found for annealed materials. This has to be seen in connection with reduced fields for saturating the magnetization, found in hysteresis measurements, and the dependence of GMR on the magnetized state of the material. The same reductions in magnetization saturation fields were found for annealed HPT-deformed materials [16].

Having a closer look at the Cu–Co-system, Berkowitz et al. [4] found higher GMR for magnetron sputtered Cu81Co19 and Cu72Co28 annealed at 484 °C, in comparison to the as-sputtered state. Giving an idea on the quantity of GMR in these studies: Measured at room temperature (RT) and in magnetic fields of 20 kOe, the resistance varies between 1% and 7% depending on material's composition and history. Using the same production route, Xiao et al. [5] found the highest GMR (16.5% measured at 5 K) when annealing Cu80Co20 for 10 min at 500 °C. Screening the GMR for different melt-spun Cu–Co-compositions by Hütten and Müller [12] yielded highest GMR at 10 K for 10 at% Co after annealing at 420 °C for 30 min. In summary, the ideal ferromagnetic content for highest GMR has to be sought for at low ferromagnetic compositions, with dedicated annealing treatments adjusted to the production route and thus to the starting microstructure.

In case of the occurrence of GMR, the drop in resistance of a magnetized material in comparison to a non-magnetized one is due to a decrease in spin-dependent scattering at interfaces of ferromagnetic and non-ferromagnetic materials. For increased granular GMR-effect, a high number of small particles (i.e. interfaces) is needed. SPD by HPT [17–19] is capable of providing these small particles, as it was already shown in [9–11]. Furthermore, it was presented in [11] that the HPT processing conditions and thus the evolving microstructure directly influence the magnitude of the drop in resistance. For example, compositions of medium ferromagnetic content provide higher GMR due to microstructural changes, when being deformed at elevated temperatures, compared to RT deformation [11]. During SPD, non-equilibrium, supersaturated, metastable phases can develop [20]. Consequently, not only coarsening but also segregation processes will take place upon subsequent annealing treatments. While growth of ferromagnetic particles will lead to a decrease in the GMR-effect, segregation of new, small ferromagnetic particles out of the supersaturated phase will increase the effect.

It has not yet been investigated whether the severely deformed state already constitutes the ideal state, i.e., maximal drop in resistance in magnetic fields, or whether there is an optimum temperature treatment to achieve this state and if so – what is it for different compositions. In this study, adequate subsequent thermal treatments for maximizing the GMR are determined and related to the underlying microstructure. Furthermore, it will be demonstrated that the SPD process by using HPT is capable of producing samples of large volume (diameter: 30 mm, height: 7 mm) with the severely deformed material exhibiting a very high microstructural stability as well as a high temperature stability of GMR properties.

## 2. Materials and Methods

Binary powder blends of Cu (Alfa Aesar, Ward Hill, MA, USA, 99.9%, -170 +400 mesh) with either Co (Goodfellow, Hamburg, Germany, 99.9%, 50–150 µm) or Fe (MaTeck, Jülich, Germany, 99.9%, -100 +200 mesh) were used for sample production. Powder blends were stored in a glove box, compressed in argon atmosphere or air and deformed to very high strains at RT using an HPT device. The applied strain linearly increases with the radius of the HPT-disc.

Samples of varying low ferromagnetic content, where according to literature highest GMR can be expected [13], were produced. On the other hand, materials of medium ferromagnetic content show saturation of magnetization in smaller fields [21,22] and consequently, saturation of GMR is achieved easier. Thus, one sample of medium ferromagnetic content was produced. While specimens of low ferromagnetic content were deformed with small HPT-tools (diameter ~8 mm, resulting sample thickness ~0.5 mm), the specimen of medium Co content was deformed with HPT-tools of 30 mm diameter (thickness ~7 mm), demonstrating the ability to upscale the HPT-process. Table 1 gives an overview of the produced samples, including their exact composition determined with energy dispersive X-ray (EDX) spectroscopy.

**Table 1.** High-pressure torsion (HPT)-processed samples, with their composition in at% and processing parameters such as HPT-tool diameter and number of rotations.

| Composition [at%] | HPT-Tool and Sample Diameter [mm] | No. of Rotations |
|---|---|---|
| $Cu_{82}Co_{18}$ | 8 | 100 |
| $Cu_{70}Co_{30}$ | 8 | 100 |
| $Cu_{55}Co_{45}$ | 30 | 250 |
| $Cu_{76}Co_{24}$ | 8 | 100 |
| $Cu_{69}Co_{31}$ | 8 | 100 |

Long and thin specimens for measuring the resistance at RT via the 4-point-probe method were prepared by careful cutting and grinding. For GMR-specimens made out of the smaller HPT-discs, the distance of GMR-specimen to the center of the disc was at least 1.5 mm, for the large HPT-disc at least 9 mm. The GMR-specimens were prepared along a secant line of the small HPT-discs, while the GMR-specimen is in axial direction for the large disc (see Figure 1 for positioning the GMR-specimen within the HPT-sample). The thickness of GMR-specimens was ~200 µm, the width ~1 mm, the length (current flow direction) is several millimeters. For measuring the GMR, the specimens were placed in the air gap of an electromagnet. RT resistivity was determined in fields up to 22.5 kOe, taking into account two different current flow orientations (parallel and perpendicular) in respect to the magnetic field. Please note, the maximum achievable magnetic field is smaller than the saturation field of most investigated materials. Consequently, the reported GMR values are not saturation values, but values measured at the maximum field of 22.5 kOe.

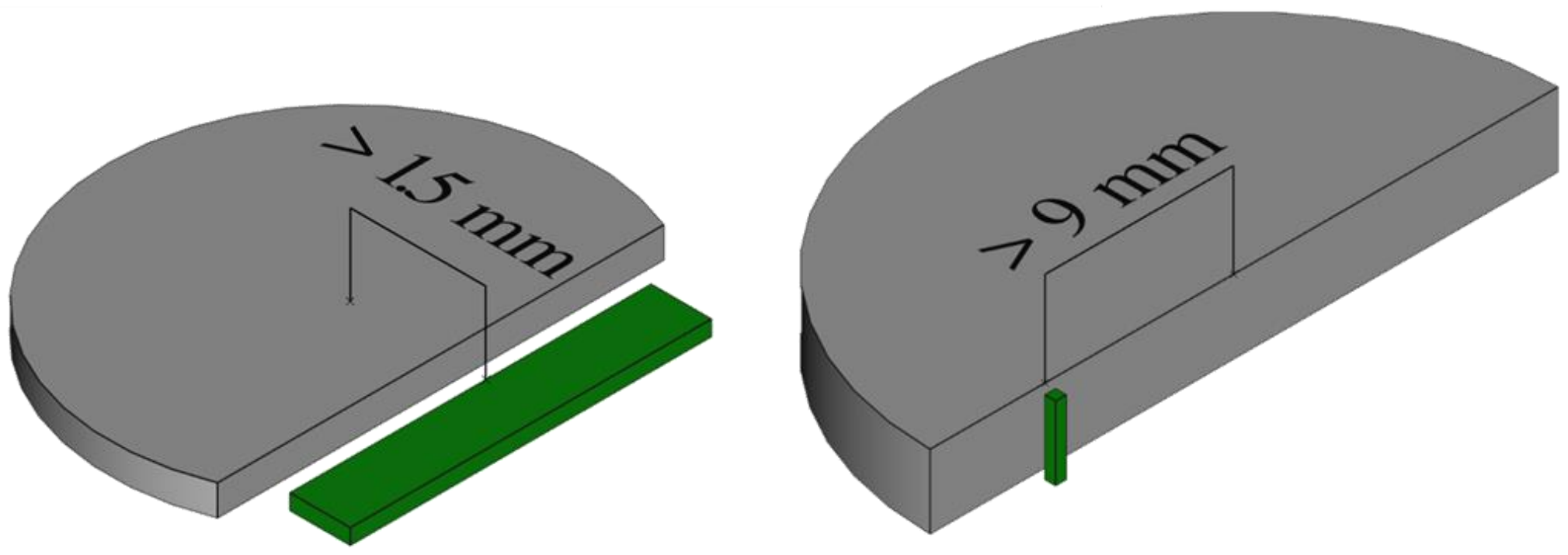

**Figure 1.** Schematic drawings of the giant magnetoresistance (GMR)—specimen position (green) within small (8 mm diameter, left) and large (30mm diameter, right) HPT-discs.

Microstructural and chemical investigations of the as-deformed, annealed HPT-samples and GMR-specimens were performed using a scanning electron microscope (SEM) equipped with EDX (SEM: LEO1525, Zeiss, Oberkochen Germany, EDX: e-Flash, Bruker, Berlin, Germany). All compositions within this publication were determined by EDX and are given in atomic percentage if not stated otherwise.

High-energy X-ray diffraction (HEXRD) experiments were performed at beamline P21.2 [23] at the synchrotron PETRA III (DESY, Hamburg, Germany) using a photon energy of 60 keV and a beam size of about 0.2 x 0.2 mm on the sample. The diffraction patterns were recorded with a Varex XRD 4343 flat panel detector and azimuthally integrated using the pyFAI software package [24]. Besides ex-situ measurements of the material in as-deformed and annealed states, the complete process of annealing was captured for one specimen by in-situ HEXRD. The sample was annealed for one hour at 490 °C and diffraction patterns were recorded every five seconds during heating, isothermal holding, and cooling. For doing so, the specimen was placed on a THMS600 heating microscope stage (Linkam, Tadworth, United Kingdom), which was flushed with argon, suppressing oxide formation. Due to thermal gradients, the thermocouple of the heating stage deviates from the sample temperature at the beam position. Thus, temperature calibration was performed via the thermal expansion of a pure copper sample that was also measured in-situ using the same conditions. Taking into account this calibration, the heating rate was found to be 9.3 K/min, the maximum temperature $T_{max}$ of 490 °C was kept for 1 h and the initial cooling rate upon furnace cooling was -120 K/min, gradually decreasing.

## 3. Results

### 3.1. In-Situ HEXRD Measurements

In order to better understand microstructural changes of the severely deformed materials upon thermal treatments, one composition out of the Cu–Co-system, Cu55Co45, was annealed in an in-situ HEXRD experiment. Figure 2 shows XRD patterns at well-defined points in time of the annealing treatment. Changes in peak shape and peak shift can easiest be seen for large diffraction angles 2θ, e.g., in the region between 11° and 12°. There, the deconvolution of the peaks into a Cu–rich and a Co-rich phase using two pseudo-Voigt functions can be expected to be more accurate. The assumption of two prevailing phases was based on the appearance of broad initial {311}- and {222}-peaks (lowest curve in Figure 2). On the other hand, Bachmaier et al. [25,26] found a single, supersaturated face centered cubic (fcc) phase, when investigating HPT-deformed material of a different chemical composition (Cu74Co26). Thus, the appearance of one or two phases in the as-deformed state can most likely be attributed to the difference in composition of the investigated materials.

The Cu–Co fcc-phases stayed stable throughout the complete in-situ annealing experiment. No complete demixing or phase transformation of the fcc-Co-rich phase into hexagonal close packed (hcp) Co has been observed after holding the temperature for 1 h at 490 °C. Due to the passing of the fcc-hcp transition temperature of Co (422 °C [27]) upon cooling, a substantial increase in hcp-Co content could be expected. However, the amount of hcp-Co after thermal treatment was found to be negligible, see uppermost pattern in Figure 2.

Comparison of the peak positions with calculated peak position for pure Cu (a = 0.3615 nm [28]) and pure $Co_{fcc}$ (a = 0.3537 nm [29]) shows that some Co is dissolved in Cu and vice versa. However, upon annealing both peaks tend towards the position for pure elements and the double peaks become more prominent. According to Vegard's law [30] and taking the mean value of {311}- and {222}-peaks, the compositions $Cu_{94}Co_6$ and $Cu_{51}Co_{49}$ are calculated for the two fcc-phases in the as-deformed state. For the material annealed at 490 °C, they change to $Cu_{97}Co_3$ and $Cu_{42}Co_{58}$. Both phases, fcc-Cu-rich and fcc-Co-rich, are still supersaturated. The occurrence of supersaturated, metastable states after HPT deformation was already found for a variety of systems, such as Cu–Ag [31], Cu–Co [32], and Cu–Fe [16].

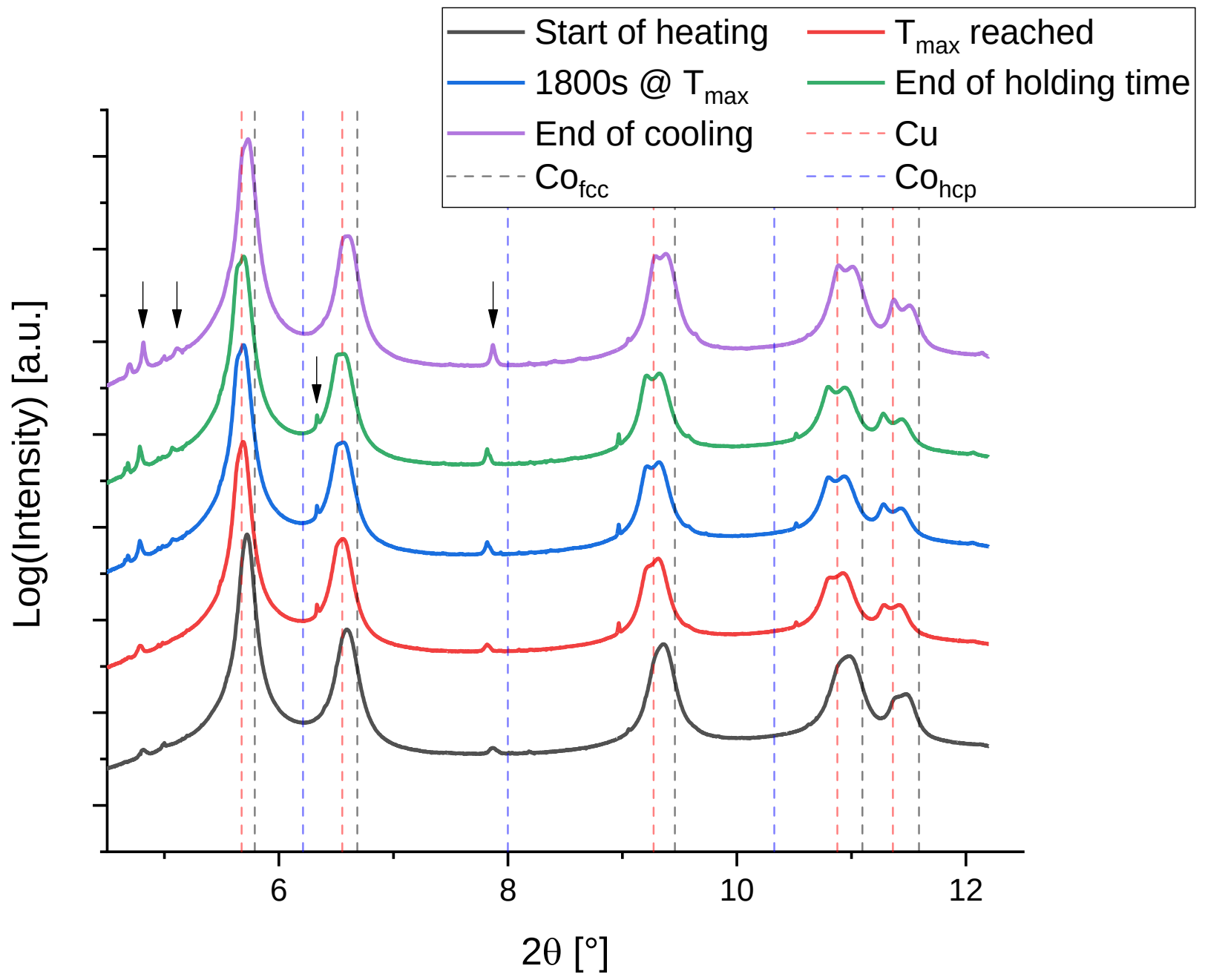

**Figure 2.** Five high-energy X-ray diffraction (HEXRD) patterns taken out of the sequence of more than 1200 patterns, recorded during the in-situ annealing treatment of Cu55Co45 at 490 °C. The time frame shown above covers the complete experiment. Gradual peak splitting and double peak formation, can easiest be seen for high diffraction angles. However, the apparent peak shift for patterns taken at elevated temperatures is mainly due to thermal expansion. Note, not all hcp-Co lines (data taken from [33]) are shown, but those where the absence of hcp-Co becomes evident. Small peaks at diffraction angles of ~5° and ~8° can be explained by oxides such as CoO, CuCoO, or Cu2O. Those that are clearly identified are marked with arrows in the uppermost patterns.

To give an even better idea of the processes taking place during the in-situ annealing, contour plots of the measured intensity are provided in Figure 3. This includes a detailed view for large 2θ-values (Figure 3b), where peak shifts and splitting in course of the experiment can easiest be followed. For doing so, pseudo-Voigt profiles are fitted in parts of each recorded pattern and the resulting center values for {222}-planes of Cu and Co are displayed in Figure 3b. The region of temperature increase, up to the time of ~3000 s, is governed by a linear decrease in peak position due to thermal expansion. Within the region of constant temperature (time: ~3000 s – ~5600 s) the {222}-peaks tend towards the positions of pure elements, due to ongoing demixing of Cu and Co. At maximum temperature, the Co-line shifts more strongly than the Cu-line. The final peak positions after cooling down, are slightly shifted towards the pure elemental lines. However, judging from the gap between fitted center peak position and the literature values for pure elements, some decomposition occurred, but both fcc phases are still supersaturated – as described above.

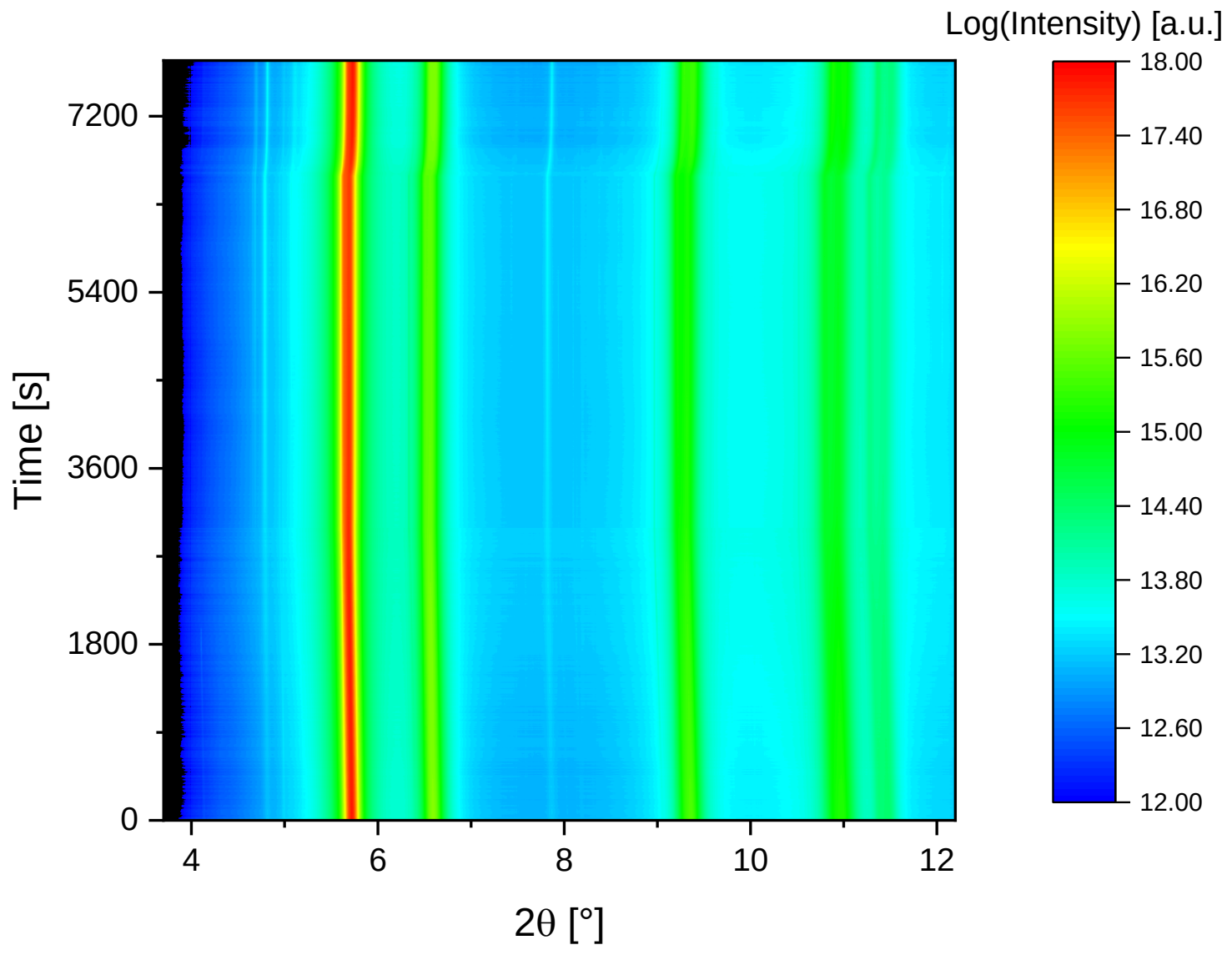

**(a)**

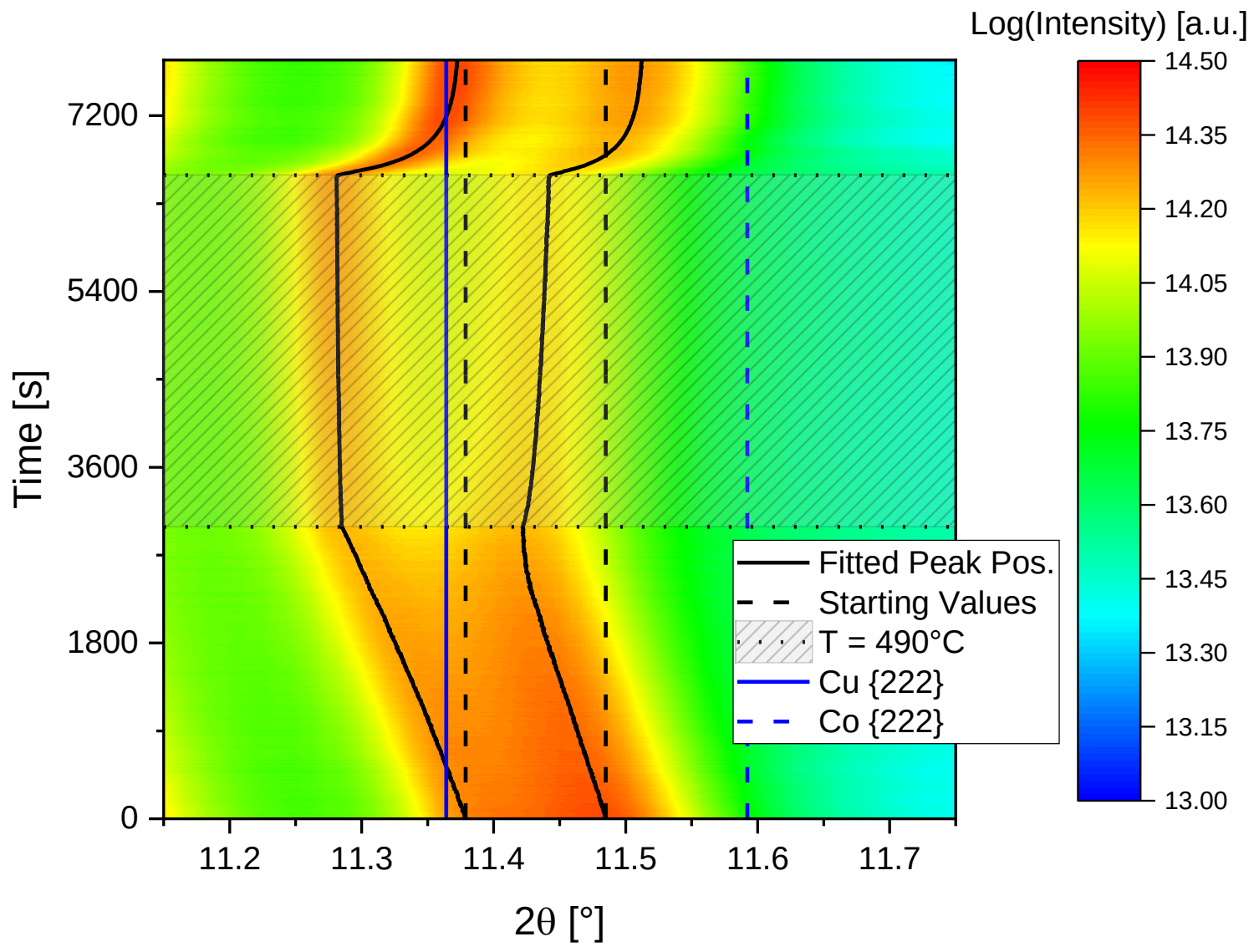

**(b)**

**Figure3. (a)** Intensity plot of the HEXRD data of the complete in-situ annealing treatment of Cu55Co45. The detailed view at high diffraction angles, (**b**), gives a better view on peak shifts as a function of applied temperature. The fitted center position of {222}-peaks of Cu and Co, including the starting values, the position of {222}-peaks for the pure elements, as well as the region of constant temperature (Tmax = 490 °C) are presented. The heating curve is provided in Figure 4.

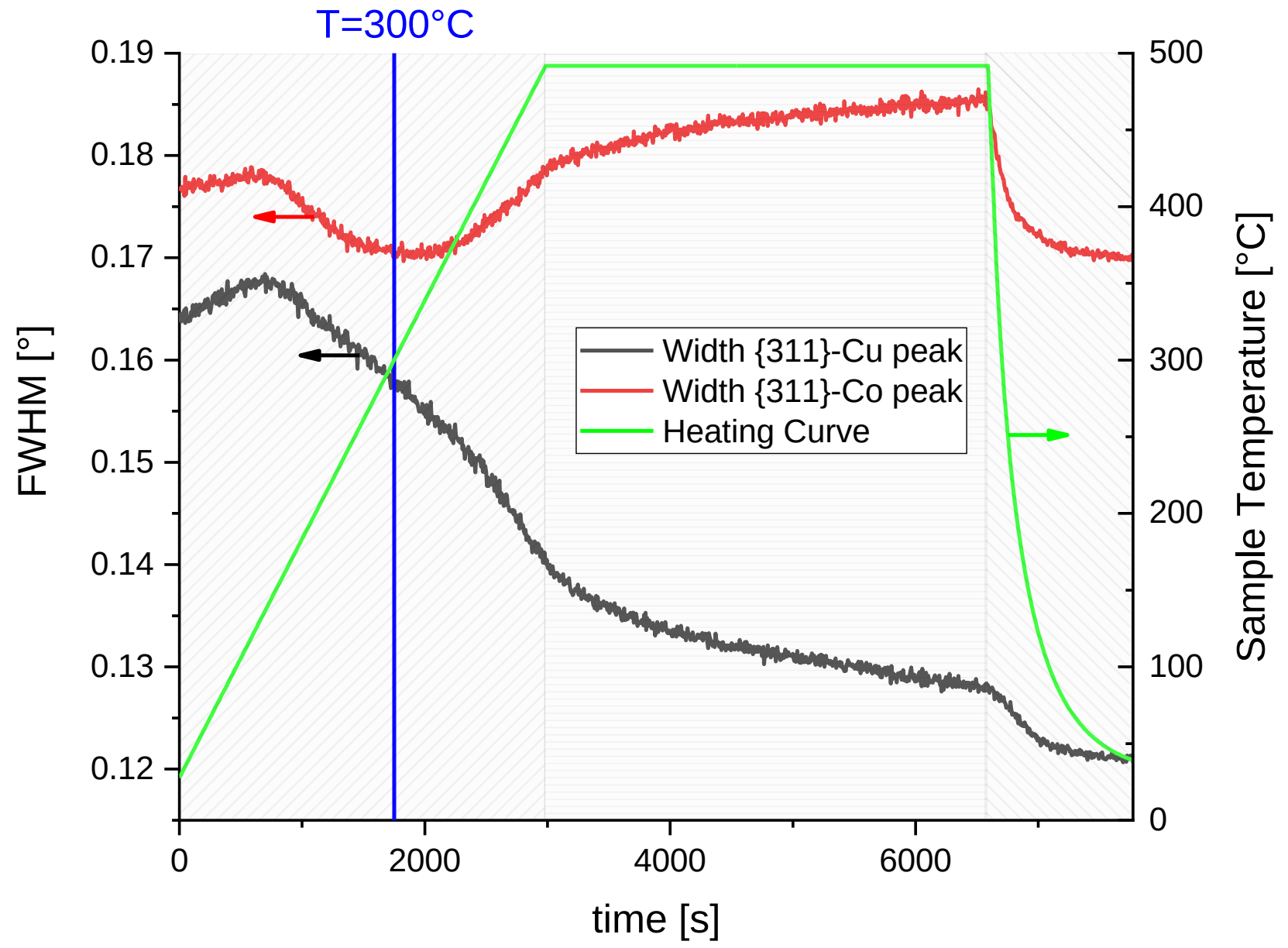

**Figure 4.** Full width at half maximum (FWHM) of pseudo-Voigt function in course of the complete in-situ HEXRD experiment. The heating curve is shown and the time when the specimen reaches a temperature of 300 °C is emphasized.

### 3.2. Ex-Situ Annealing Treatments

Recapturing from the in-situ experiments: The initial state of the specimen used for the in-situ HEXRD experiment (Figure 2, lowest pattern) shows the co-existence of two fcc phases, enriched in Cu and Co. During the course of annealing, the faint double peaks (e.g., the one for Cu- and Co-{222}, Figure 3b) become better visible due to peaks becoming sharper and being further in distance. Having a closer look at the width of {311}-peaks (Figure 4), a resultant of fitting pseudo-Voigt functions, the expected trend of decreasing width for increasing Cu grain sizes can be seen. For the {311}-peak of Co, this trend is not alike the one of {311}-Cu for temperatures above 300 °C.

One explanation of increasing peak width might be newly formed, small Co-particles. Consequently, the maximum GMR was searched for in this region. The annealing temperatures of 150 °C, 300 °C, and 400 °C were investigated. This was done in combination with an annealing treatment at 600 °C, where well-advanced decomposition of Co and Cu can be expected [26]. The length of all annealing treatments was 1 h. While the annealing treatment at 150 °C was performed in air followed by quick cooling, all other annealing treatments were performed in vacuum with subsequent slow furnace cooling. Figure 5 shows ex-situ HEXRD patterns after these annealing treatments together with the last profile, after cooling, of the in-situ annealed specimen. Up to 490 °C weak but gradual separation of the peaks can be seen. As the peak positions for the material annealed at 600 °C coincide well with values reported in literature, this points towards a full separation of phases.

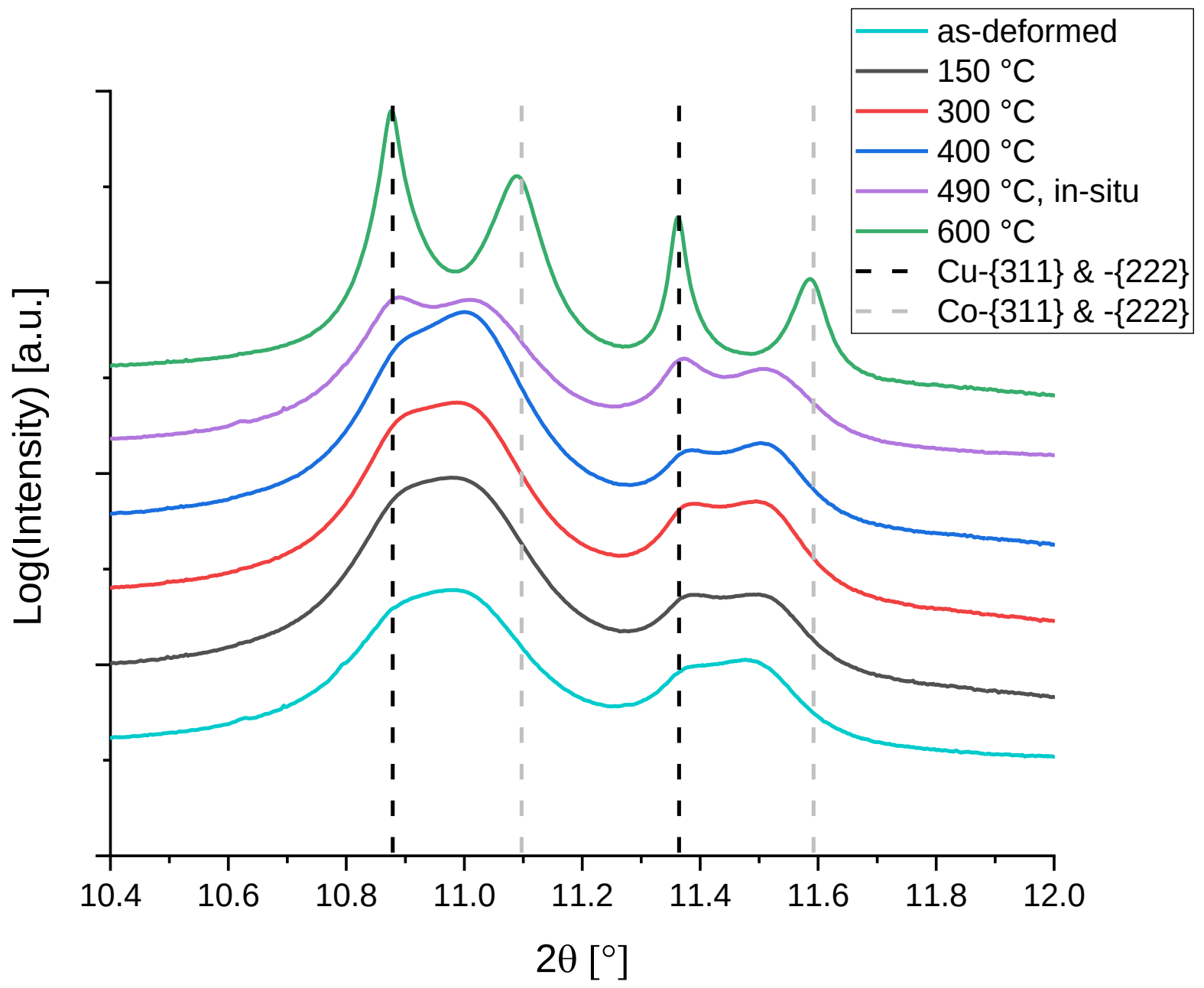

**Figure 5.** HEXRD pattern of the as-deformed and ex-situ and in-situ annealed Cu55Co45 specimens, in addition the lines of pure elements are shown.

The microstructure of the as-deformed and annealed specimens was additionally investigated using a SEM (Figure 6). The coarsening of the initially nanocrystalline microstructure with higher temperatures is evident. Co-particles, appearing darker in backscatter mode due to Z-contrast, can be identified with SEM after annealing at 400 °C for the first time. The sample, which has been annealed at 600 °C shows a high phase contrast, again indicating a separation of phases but only slight concomitant grain growth can be observed. The high temperature stability of the microstructure up to temperatures of 600 °C, more than 0.6 of the homologous temperature of copper, is remarkable. This high temperature stability is typical for some nanocrystalline materials processed by HPT. In [34], an even higher thermal stability (0.7 of homologous temperature) was found for HPT-deformed Ag reinforced with nanodiamonds.

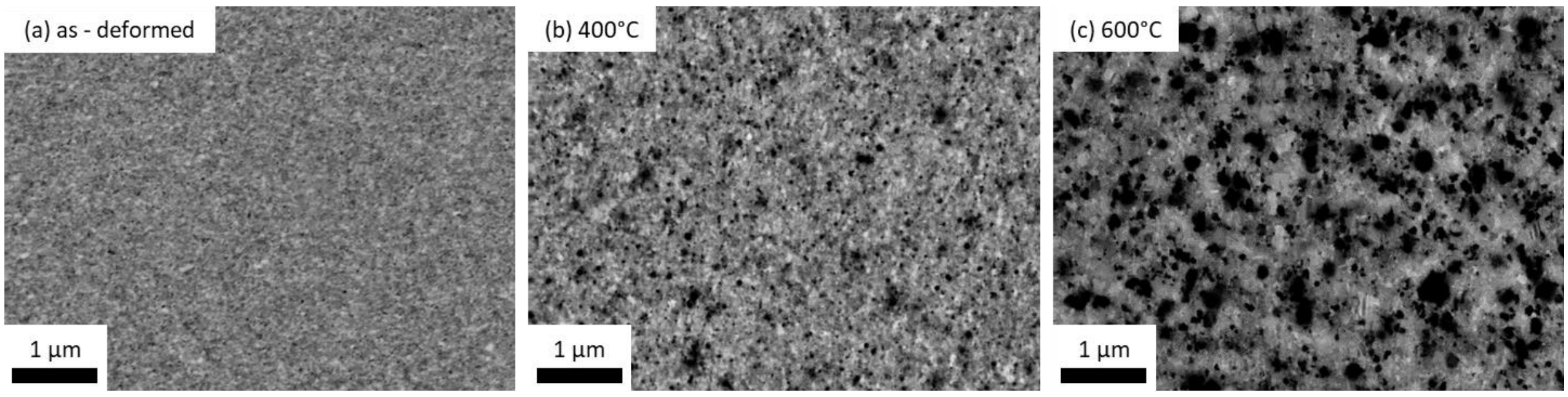

**Figure 6.** Micrographs of Cu55Co45 in three different microstructural states in tangential view, as-deformed (**a**) and annealed for 1 h at 400 °C (**b**) and 600 °C (**c**).

### 3.3. Medium Ferromagnetic Content: Magnetoresistance Measurements

Specimens for GMR-measurements were prepared out of the as-deformed and the annealed Cu55Co45 material and the drop in RT resistivity was recorded as a function of applied magnetic field. Furthermore, the specimens were placed i) with the direction of applied current perpendicular to the magnetic field and ii) with the current parallel to the magnetic field. Specimens using the same production route and of varying Co-contents were already investigated under the same conditions and the results are presented in [11]. Therein, a gradual change from isotropic GMR for low ferromagnetic content to AMR with increasing Co-content is reported. Figure 7 shows that the GMR is not perfectly isotropic for the Cu55Co45, owing to the comparatively high Co-content, indicating some contribution of AMR. Parallel alignment of current flow with the magnetic field gives slightly smaller drops in resistance.

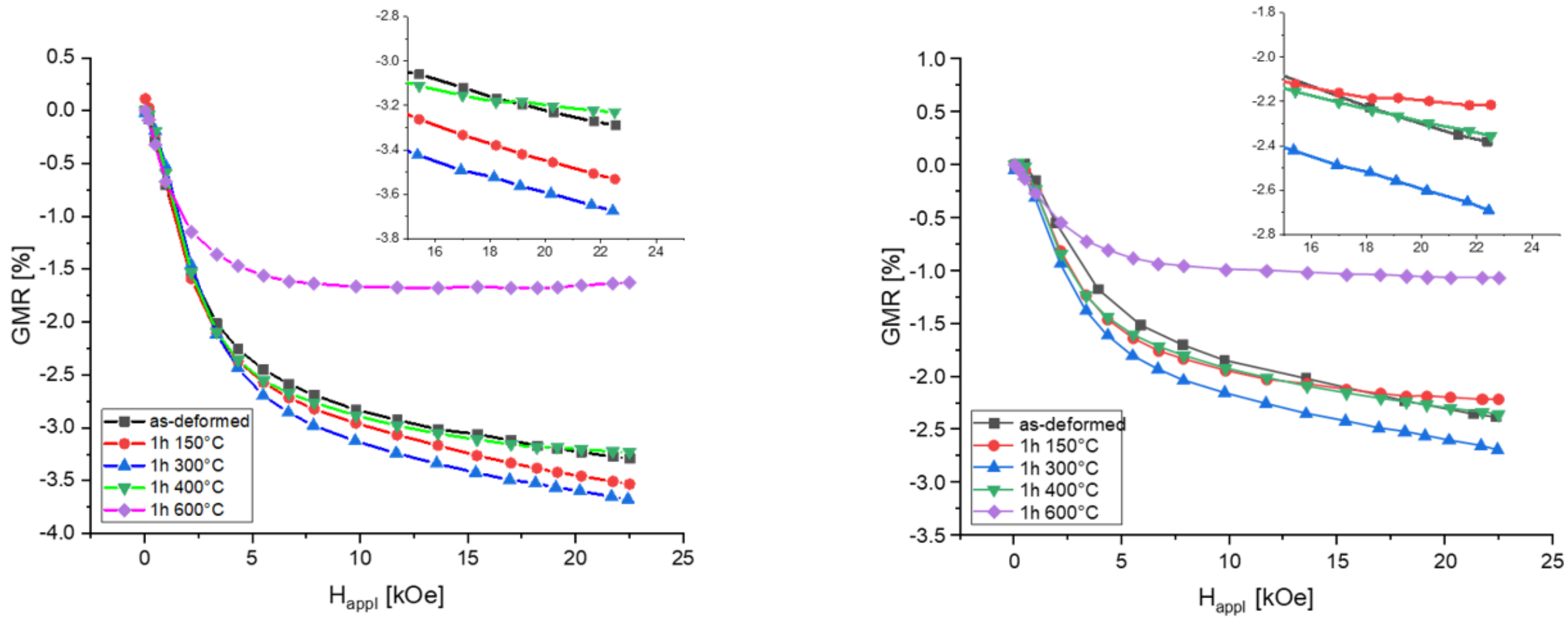

(a) (b)

**Figure 7.** Drop in resistance as a function of applied magnetic field for Cu55Co45 in as-deformed and different annealed states. (**a**) Current applied perpendicular to the magnetic field, (**b**) current applied parallel to the magnetic field.

Figure 7 depicts the highest GMR-effect for the specimen annealed at 300 °C, in both directions. Except for the 150 °C annealed specimen in parallel orientation, there is the general trend of increasing GMR-effect up to 300 °C followed by a decrease. As expected, the 600 °C annealed specimen shows minimal GMR, due to the large Co-grains. Furthermore, its saturation occurs at smaller magnetic fields, pointing at an enhanced susceptibility.

### 3.4. Improving the Magnetoresistive Effect

After screening the influence of annealing temperature onto the GMR-behavior of a material of medium ferromagnetic content, the same treatment is performed on materials of lower ferromagnetic content. As it is known from literature [13], there is an optimum ferromagnetic content of approximately from 15 to 25%. The HPT-processability for binary Cu–Fe and Cu–Co alloys was investigated in [21], where it was stated that samples of low Co content feature large remaining Co-particles, while the deformation of samples of increasing Fe-content (> ~25%) is accompanied with shear band formation. Furthermore, preliminary experiments on annealed Cu85Fe15 showed the limitations of the existing setup, due to the specimen's increased conductivity and the small RT GMR-effect (~0.2% [11]). Details on investigated sample composition can be found in Table 1.

The HEXRD results, shown in Figure 8 demonstrate the supersaturation of Cu with Co or Fe, except for the specimen of Cu69Fe31 still showing a substantial amount of the body centered cubic (bcc) phase.

GMR was determined for the specimens in the as-deformed and in one annealed state. An annealing temperature of 300 °C yielded best results for Cu55Co45. Thus, this temperature was chosen to be used for annealing of MR-specimens of low ferromagnetic content and the results are summarized in Figure 9. Therein, only the perpendicular specimen orientation is shown, due to the almost perfect isotropic behavior. GMR increases for the annealed Cu–Co, while the 300 °C-annealed Cu–Fe already decreases in GMR. The occurrence of the maximum GMR already in the as-deformed (or as-sputtered state, as it was the case for Wang and Xiao [5]) state cannot be ruled out. Decreasing GMR-effect for annealed Cu–Fe can be explained by formation and growth of Fe grains at low temperatures. HPT-deformed Cu–Fe already shows an increasing bcc-Fe fraction when annealing at 150 °C [16] and substantial decomposition at 500 °C [16,22], whereas Cu–Co is stable regarding substantial decomposition up to at least 490 °C (this work, for higher Co-content [36]). For both Fe-containing specimens, GMR is higher in comparison to Cu85Fe15 [11] with a RT GMR in the as-deformed state of 0.2%.

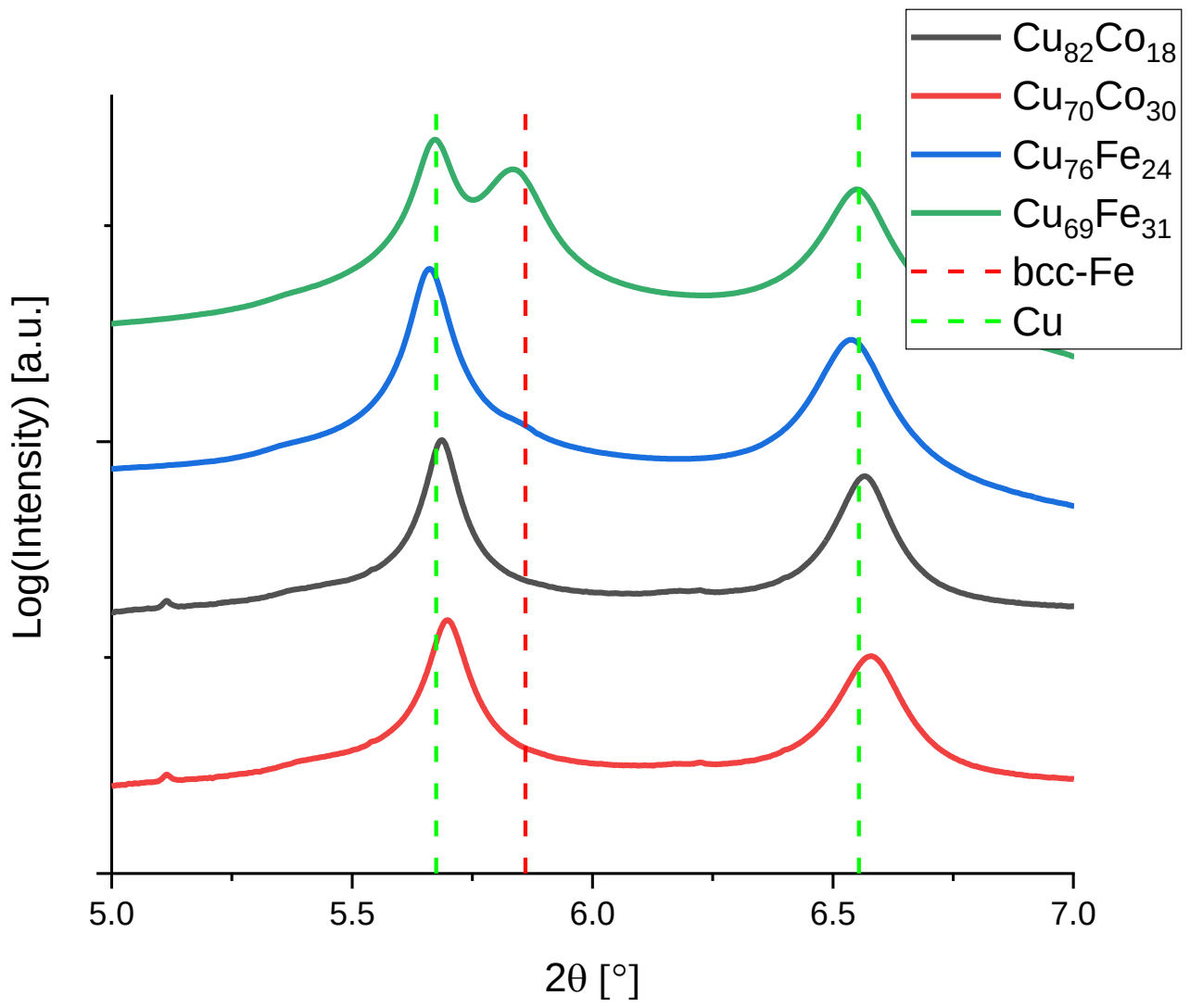

**Figure 8.** HEXRD pattern of Cu–Co and Cu–Fe samples of low ferromagnetic content, in as-deformed states. In addition, the lines of pure elements are shown, data for bcc-Fe was taken from [35].

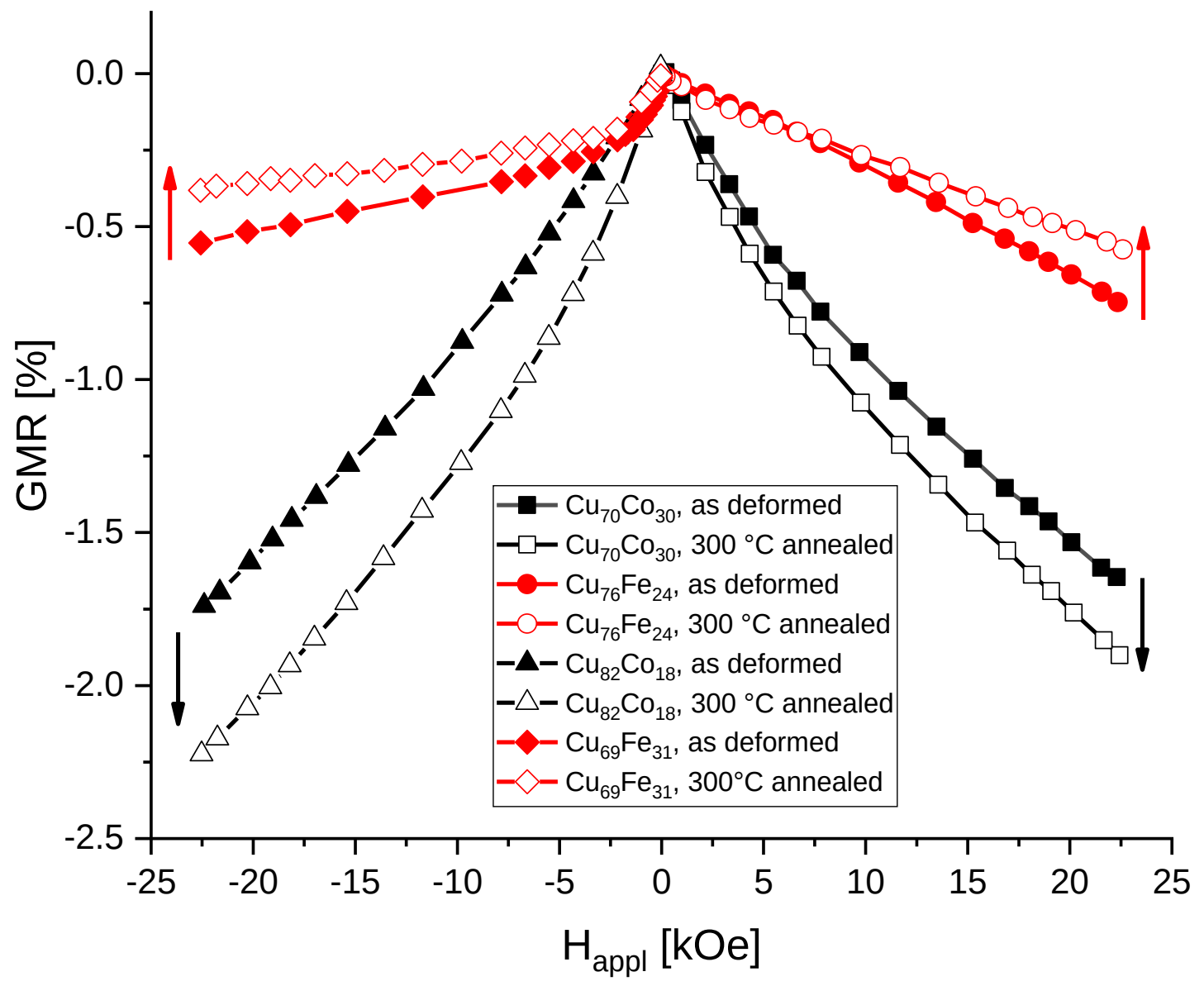

**Figure 9.** GMR as a function of applied magnetic field for specimens of low ferromagnetic content in the as-deformed state and after annealing for 1 h at 300 °C. Current flow was perpendicular to the applied magnetic field.

A summary of all GMR values in magnetic fields of 22.5 kOe can be found in Figure 10.

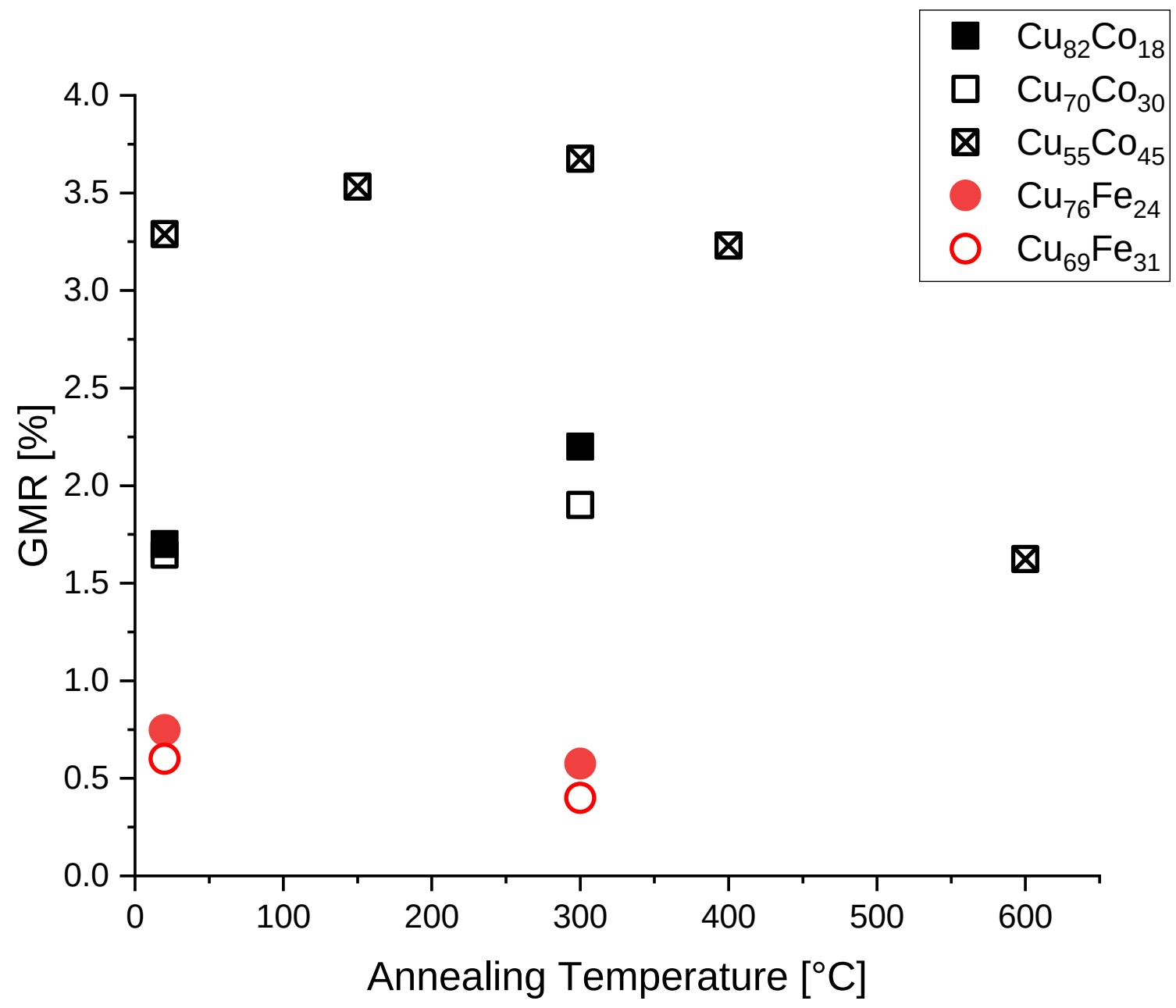

**Figure 10.** Absolute value of GMR for all investigated Cu–Co and Cu–Fe materials and for different annealing temperatures. Values are measured in magnetic fields of 22.5 kOe. While Cu55Co45 shows highest GMR for annealing at 300 °C and all other Co-materials also increase in GMR, this is not the case for Cu–Fe, where a smaller GMR was found after annealing at 300 °C.

## 4. Discussion

### 4.1. Is the Change in Resistance a Magnetic Effect?

Indeed, there is an improved GMR-effect upon annealing in Cu–Co. The RT relative resistance at high magnetic fields – in relation to the resistance at zero field – is smallest for the Cu55Co45 annealed at 300 °C. The same tendency holds true for smaller Co contents. Please note, the higher GMR-effect found for the Cu–Co sample of medium composition is due to the achievable magnetic field not being high enough to saturate material of low ferromagnetic content.

For the determination of GMR, the change in resistance within the magnetic field is set in relation to the resistance at zero magnetic field (Equation (1)). Thus, the improved GMR-effect could also be due to a lower overall resistivity upon annealing. Miranda et al. [37] tried to disentangle magnetic contributions and contributions from microstructural changes upon annealing (e.g., from stress relief, recovery processes). This was done by employing Equation (2)

$$\frac{\rho_{mag}(H)}{\rho_0} = \frac{\rho(H) - \rho(H=0)}{\rho(H=0)}\left(1 + \frac{\Delta\rho_T}{\rho_0}\right) \quad (2)$$

uncovering the magnetic contribution to the GMR. Therein, $\rho_0$ is the resistivity before annealing, the first factor on the right side the typical form for deriving GMR (see Equation (1)) and the second factor containing the relative change in resistivity upon thermal treatment at the temperature T. Application of Equation (2) on the experimental data yields the same behavior for the magnetic contribution as it was found for the GMR itself. Namely an increase with increasing annealing temperature, followed by a drop in $\rho_{mag}$ / $\rho_0$ above 300 °C. Thus, as it was also judged in [37], this behavior can mainly be associated with a magnetic effect.

### 4.2. Where Does the Change in GMR Come From?

The starting material for the in-situ synchrotron experiment shows Cu-rich and Co-rich fcc-phases (see lowest pattern in Figures 2 and 5), the amount of hcp-Co is negligible. Having a closer look at the width of the fitted pseudo-Voigt functions, a different evolution of the Cu- and the Co-peak of {311}-planes (Figure 4) is found. The same qualitative behavior was also found for {222}-planes. While for the Cu-peak, the expected decrease in peak width due to recovery and coarsening is found, the width of the Co-peak increases, starting at approximately 300 °C. As XRD is more sensitive to small particles compared to, e.g., SEM, one explanation of increasing width might be the formation of small Co-particles out of the Cu-rich phase. In this case, the resulting Co-peak is a superposition of the small particles' peak and the one of the original Co-rich phase (possibly showing the same trend as the Cu-line does). The change of the lattice parameter of the Cu-rich phase due to Co-depletion is small (see Figure 3b), nonetheless existent and an additional amount of additional non-ferro and ferromagnetic interfaces would increase the GMR-effect. Most importantly, although there is a coincidence of temperature yielding maximum GMR and a change in the evolution of the peak width of the ferromagnetic phase, further transmission electron microscopy and atom probe microscopy studies are a prerequisite to finally confirm the exact causes of this result.

As it was already stated, the saturation of GMR occurs for Cu55Co45 annealed at 600 °C (1.5% at 22.5 kOe), already within the window of achievable magnetic fields. Thus, another aspect of thermal treatments, the easier saturation of the magnetization and consequently of the GMR has to be considered. For judging small changes in resultant GMR at a certain field, this becomes especially important when the difference in GMR upon annealing is small. Hysteresis measurements of HPT-deformed Cu–Fe materials, with Fe-content below 25 wt% showed that the RT hysteresis does not saturate in fields up to 7 T and superparamagnetic states were found for Cu86Fe14 and Cu75Fe25 (in wt%) [21]. The same behavior of non-zero high field susceptibility was found for Cu72Co28 (in wt%) [13], where Wang and Xiao stated that the occurrence of superparamagnetism does not influence the saturation value of GMR but its saturation process. Consequently, small particles are preferred for achieving high GMR, but thermal treatments have the advantage of reducing the saturation fields. As the saturation value of GMR was not measurable for all but the 600 °C annealed material, measurements of the saturation values of GMR would have to be made at very high fields.

### 4.3. What Is the Optimum Composition and Thermal Treatment after RT HPT?

Regarding the optimum composition for maximum GMR, the results of this study shall be discussed in combination with results taken from [11]. Therein, the RT GMR of both, Cu81Co19 and Cu64Co36 was approximately 1.5% in the as-deformed state, while the one of as-deformed Cu70Co30 is slightly above. Thus, a compositional region in between ~20 at% and ~35 at% Co has to be searched for optimum GMR using smaller increments. As a reminder, these GMR values are valid for fields of 22.5 kOe—the maximal attainable magnetic field with the used equipment. The GMR of Cu55Co45 was found to be higher (~3%) at this magnetic field. However, considering literature on optimum ferromagnetic composition [13], the higher GMR-value is simply due to faster saturation.

For Cu–Fe, while the GMR-effect of Cu85Fe15 is rather small (0.2% [11]), it increases when adding Fe, see Figure 9, and an optimum composition has to be searched for above 15 at% Fe and below the limit of RT HPT-processability. Changing the HPT process parameters, e.g., the deformation temperature opens a wider field of achievable, homogeneously deformed compositions [36]. For comparison, Wang and Xiao [13] found maximum GMR in the magnetron sputtered state for Cu80Fe20.

Regarding the aspect of the absolute value of maximum achievable GMR in higher magnetic fields, it is of importance that RT HPT-processed Cu–Co-materials (<40wt%) and Cu–Fe-materials do not show a saturation of magnetization in fields of 7 T [21] and so it will not be the case for GMR. Miranda et al. [37] measured ~7% GMR for as-melt spun ribbons of Cu90Co10 at approximately 20 kOe, while measuring more than 10% at fields above 60 kOe; the value still was not saturating. Thus, it can be expected for HPT-deformed materials – when considering hysteresis loops from literature [21] – that much higher GMR values are to be expected, when the resistance measurements is performed in higher fields.

In the aspect of optimum annealing temperature, 300 °C were found to be best for medium Co-content. For small Co-content, phase and particle evolution could be slightly different but the optimum temperature is expected not to be far off 300 °C. For the Cu–Fe system, temperatures below 300 °C or a shorter duration of annealing treatment should be considered and will be investigated in the future.

## 5. Conclusion

Using SEM and synchrotron diffraction in combination with resistivity measurements in magnetic fields, it was possible to establish a link between the microstructure and functional properties of severe plastically deformed materials. Magnetoresistance was found to be improvable by adequate thermal treatments of as-deformed samples.

Different binary compositions containing Cu and Fe or Co were severe plastically deformed and investigated in the as-deformed and different annealed states. For $Cu_{55}Co_{45}$, an annealing treatment for 1 h at 300 °C was found to give the highest GMR at fields of 22.5 kOe. An increase in GMR was also found for lower Co-content using the same thermal treatment. Cu–Fe shows smaller drops in resistivity in fields of 22.5 kOe compared to Cu–Co. Furthermore, annealing Cu–Fe at 300 °C was found to be too high for improving GMR.

In general, with the technique of SPD using HPT, it is possible to produce bulk nanocrystalline materials showing GMR behavior at RT. The GMR-ratio is smaller in comparison to well-established multilayer systems, already in application, however comparable to literature values of materials showing granular GMR. The usage of SPD by HPT offers several advantages: First, the sample size processed in this paper reaches 5 $cm^3$, with a possibility to achieve sample volumes up to 34 $cm^3$ [38]. Second, for a particle size aspect ratio close to 1 of a dilute ferromagnetic phase within the large HPT sample (diameter >30 mm), an isotropic behavior independent of GMR-probe extraction direction can be expected. Third, the temperature stability of the investigated Cu–Co system is astonishingly high with the GMR-ratio remaining stable for annealing treatments for 1 h at temperatures up to 400 °C.

**Author Contributions:** Conceptualization, S.W. and A.B.; methodology, S.W., M.S., L.W., T.M., and A.B.; software, S.W. and T.M.; validation, S.W.; formal analysis, S.W.; investigation, S.W., M.S., L.W., T.M., and A.B.; resources, T.M. and A.B.; data curation, S.W.; writing—original draft preparation, S.W.; writing—review and editing, S.W., M.S., L.W., T.M., and A.B.; visualization, S.W.; supervision, A.B.; project administration, A.B.; funding acquisition, A.B. All authors have read and agreed to the published version of the manuscript.

**Funding:** This project has received funding from the European Research Council (ERC) under the European Union's Horizon 2020 research and innovation programme (Grant No. 757333).

**Acknowledgments:** S.W. deeply appreciates the help of Mirjam Spuller and Alexander Paulischin for performing the GMR specimen preparation and assistance in resistance measurements. We acknowledge DESY (Hamburg, Germany), a member of the Helmholtz Association HGF, for the provision of experimental facilities. Parts of this research were carried out at beamline P21.2 at PETRA III under proposal I-20190577.

**Conflicts of Interest:** The authors declare no conflict of interest. The funders had no role in the design of the study; in the collection, analyses, or interpretation of data; in the writing of the manuscript, or in the decision to publish the results.